\documentclass[twocolumn]{svjour3}         
\smartqed  
\usepackage{graphicx}
\begin{document}

\title{Sulfur chemistry in the Horsehead
}
\subtitle{An interferometric view of the Horsehead PDR}


\author{J.R. Goicoechea$^1$
\and J. Pety$^{1,2}$
\and M. Gerin$^1$ 
\and \\D. Teyssier$^3$
\and E. Roueff$^4$
\and P. Hily-Blant$^2$
}


\institute{$^1$LERMA--LRA, UMR 8112, CNRS, Observatoire de Paris and Ecole Normale
Sup\'erieure, 24 Rue Lhomond, 75231 Paris, France.
\email{javier@lra.ens.fr}\\
$^2$IRAM, 300 rue de la Piscine, 38406 Grenoble cedex, France.\\
$^3$European Space Astronomy Centre, Urb. Villafranca del Castillo, P.O.
Box 50727, Madrid 28080, Spain.\\
$^4$LUTH UMR 8102, CNRS and Observatoire de Paris, Place J. Janssen
92195 Meudon cedex, France.}

\date{to appear in Astrophysics and Space Science, 
special issue of "\textit{Science with ALMA: a new era for Astrophysics}" conference}

\maketitle

\begin{abstract}
Sulfur is an abundant element which remains undepleted in diffuse interstellar gas (A$_V<1$)
but it is traditionally assumed to deplete on dust
grains at higher densities and larger A$_V$.

Photodissociation regions (PDRs) are an interesting intermediate medium between 
translucent and dark clouds where the chemistry and energetics are dominated by the 
illuminating FUV radiation field. Thus they can provide  new insights about
the sulfur depletion problem. However, physical and chemical gradients in PDRs
take place at small angular scales ($\sim$1 to 10$''$). Aperture synthesis observations 
are therefore required to resolve such gradients.
Besides, a complete understanding of molecular excitation is needed to correctly determine
molecular abundances but also the preavailing physical conditions. Hence, multi-$J$ 
observations at increasing frequencies are also required.
Such high angular resolution and broad frequency coverage observations will be
provided by ALMA in the near future.

In this work we present IRAM-PdBI  observations of the CS $J$=2--1  line toward 
the Horsehead PDR complemented with IRAM--30m 
observations of several rotational lines of different
sulfur bearing molecules (CS, HCS$^+$, SO, H$_2$S, etc.).
Photochemical and nonlocal, non-LTE radiative transfer models
adapted to the Horsehead  geometry have been developed.
The gas phase sulfur abundance has been inferred in the PDR.

\end{abstract}

\section{Introduction}
\label{intro}

Sulfur is an abundant element (the solar photosphere abundance is 
S/H=1.38$\times$10$^{-5}$), which remains undepleted in  diffuse interstellar 
gas  and H\,{\sc ii} regions  but it is traditionally assumed
to deplete on grains in higher density molecular clouds by factors as large as $\sim$1000. 
This conclusion is \textit{simply} reached by adding up the observed gas 
phase abundances of S--bearing molecules in well known dark clouds such as TMC1.
  As sulfur is easily ionized  (ionization potential $\sim$10.36~eV), 
sulfur ions are probably the dominant gas--phase sulfur species in translucent gas. 
Ruffle et al. (1999) proposed that if dust grains are typically negatively charged, S$^+$ may 
freeze--out onto dust grains during cloud collapse more efficiently than neutral species
such as oxygen. However, the nature of sulphur on dust  grains (either in mantles or 
cores) is not obvious. 
The absence of strong IR features due to S--bearing
ices in many ISO's mid--IR spectra  and the 
presence of S\,{\sc ii} recombination lines in dark clouds such as Rho Ophiuchi 
 all argue against a large depletion of sulfur from the gas phase.
Besides, the abundance of species such as CS or H$_2$S may indicate that something 
important is lacking from chemical models or that an abundant sulfur--bearing carrier
has been missed.  Therefore, the abundances of sulfur species remain interesting puzzles 
for interstellar chemistry.
PDRs offer an ideal intermediate medium between diffuse and dark cloud gas to investigate 
the sulfur depletion problem. In particular, 
the Horsehead western edge is a PDR viewed nearly edge-on. 
The intensity  of the incident FUV radiation field is $\chi\simeq60$ relative to the 
interstellar radiation field (ISRF) in Draine's units.
At a distance of 400 pc, the Horsehead PDR is very close to 
the prototypical source needed to serve as a reference to models (Pety et al. 2006).

\begin{figure*}[t]
  \includegraphics[width=0.9\textwidth]{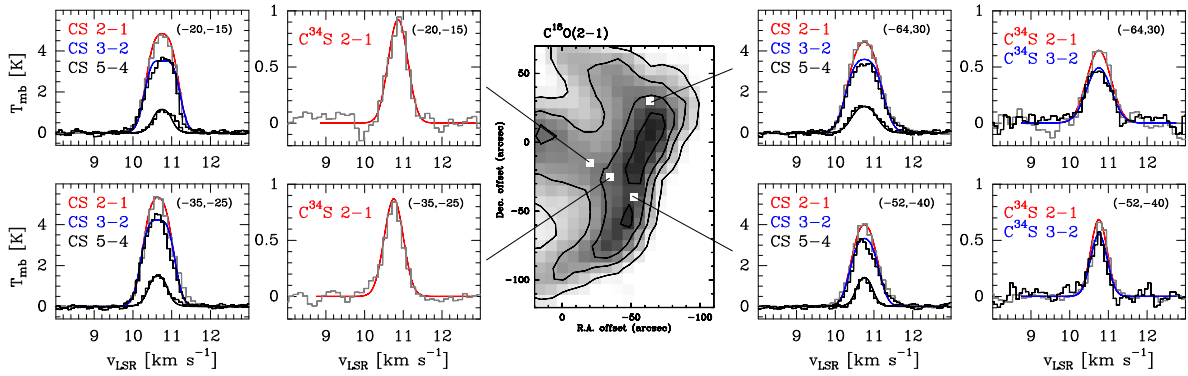}
\caption{Monte Carlo radiative transfer models for CS and C$^{34}$S discussed in the text  (curves)
that best fits the IRAM--30m observations (histograms).
Offsets in arcsec refer to the (0,0) position of the C$^{18}$O(2--1) map (middle).
Predicted line profiles have been convolved 
with the telescope angular resolution.
Intensity scale is in main beam temperature.}
\label{fig:1}       
\end{figure*}

Since 2001 we have been studying the Horsehead nebula at high angular resolution ($<5''$)
with the  IRAM PdBI interferometer, and complemented with larger scale and higher
frequency observations with the   IRAM-30m single--dish telescope. 
Following these and other works, the Horsehead PDR is now considered a well understood source,
ideal to analyse the chemical stratification
predicted in clouds illuminated by a FUV radiation field
(see Pety et al. these proceedings). These studies have led e.g., to a significant increase
of our knowledge of the carbon chemistry (from PAHs to small hydrocarbons)
in FUV irradiated gas (see Gerin et al. 2005; Pety et al. 2005).
More recently,  we have tried to accurately determine and understand the observed
CS and HCS$^+$ abundances 
in the Horsehead PDR as a tool for estimating  the sulfur gas phase abundance.
Compared to other works, we now specifically treat the nonlocal, non-LTE
molecular excitation and radiative transfer along the cloud edge using the output predictions
of photochemical models. In this way, we can consistently link the thermal and 
chemical predictions of PDR models with the absolute molecular line intensities detected
by telescopes. This, sometimes difficult, step has a crucial importance in the
accurate determination of abundances and in our
understanding of the prevailing  chemistry and physical conditions by  a direct comparison
of PDR models with molecular line emissivities.

In this work we present  $3.65'' \times 3.34''$ angular resolution 
IRAM \textit{Plateau de Bure Interferometer}  observations of the CS $J$=2--1
line (Fig.~2~\textit{left}), 
complemented with IRAM--30m observations of several rotational lines of CS, 
C$^{34}$S, HCS$^+$, SO,  $^{34}$SO, H$_2$S, etc.
(some of them shown in Fig.~1).
Beam sizes of single--dish observations range from $\sim$10$''$ at 1~mm to 
$\sim$30$''$ at 3~mm.
Data reduction was done with the IRAM GILDAS software.

\section{Numerical methodology}

To analyse the gas phase sulfur chemistry we use the \textit{Meudon PDR code},
a photochemical model of a unidimensional  stationary PDR (Le Petit et al. 2006). In few words,  
the PDR code solves the FUV radiative transfer in an absorbing 
and diffusing  medium of gas and dust (Goicoechea \& Le Bourlot 2007). 
This allows the explicit computation of the 
FUV radiation field (continuum+lines) and therefore, the explicit integration of 
consistent  C and S photoionization  rates together with H$_2$, CO, $^{13}$CO, and 
C$^{18}$O photodissociation rates.
Our standard conditions for the model of the Horsehead PDR 
include a power--law  density profile  and a FUV radiation field enhanced by
a factor $\chi=60$ with respect  to the Draine ISRF.
Different sulfur gas phase abundances, S/H, have been investigated (see  Fig.~2~\textit{right}).
To be consistent with  PdBI CO multi-line observations,
the thermal balance was solved until the gas temperature
reached a minimum value of 30~K, then a constant temperature was assumed.

For the molecular excitation and radiative transfer in the PDR we use a 
nonlocal, non--LTE Monte Carlo code that allows us to directly compare
our millimeter line observations. The code handles spherical and 
plane-parallel geometries and accounts for line trapping, collisional 
excitation, and radiative excitation by absorption of microwave cosmic 
background and dust continuum photons. Arbitrary density, temperature
or abundance profiles, and complex velocity gradients can be included.
A detailed description of the model is given in Goicoechea (2003) and
Goicoechea et al. (2006).

\begin{figure*}[th]
  \includegraphics[width=0.9\textwidth]{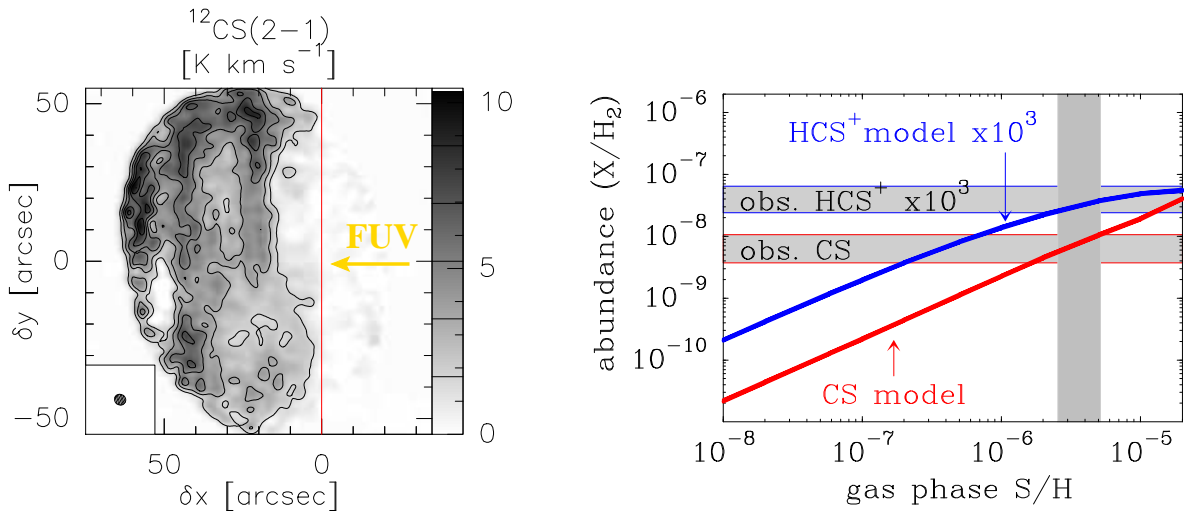}
\caption{\textit{Left panel:} CS $J$=2--1 integrated emission map obtained with the PdBI. 
The map center has been set to the mosaic
phase center: RA(2000) = 05h40m54.27s, Dec(2000) =
-02$^o$28$'$00$''$.  The map size is $110'' \times 110''$.
 The synthesized beam is plotted in the bottom left corner.
The map has been  rotated by
14$^o$ counter--clockwise around the image center to bring the
exciting star direction in the horizontal direction as this eases the
comparison of the PDR models.
\textit{Right panel:}
Photochemical model predictions for the physical and 
FUV illuminating conditions prevailing in the Horsehead PDR showing 
the CS and HCS$^+$ abundances as a function of the sulfur gas phase 
abundance. Horizontal shaded regions show the CS and HCS$^+$ abundances 
derived from the single--dish observations and radiative transfer modeling.
Note that for clarity HCS$^+$ abundances have been multiplied by a 
factor of 1000. The shaded vertical region shows the estimated sulfur 
abundance in the  Horsehead nebula derived from the constrained fits
of CS and HCS$^+$ abundances.}
\label{fig:2}       
\end{figure*}

The following methodology was carried out: a full PDR model with
Horsehead standard conditions 
was run with a particular choice of the density gradient. 
Afterwards, the PDR output was used as input
for the excitation and radiative transfer calculation. 
In this way, physical parameters can be tuned more accurately
by iteration of different radiative transfer models. 
Once better parameters have been found, a new PDR computation is 
performed with this choice of physical parameters.
We have so far analyzed CS,
C$^{34}$S, C$^{18}$O and HCS$^+$ species.
Synthetic abundance
profiles are consistently computed as a function of the edge distance
$\delta x$ (in arcsec) and directly compared with interferometric
observations (see Fig.~3). 
We are now analyzing a larger collection of sulfur bearing molecules.

\section{Main Results}

We have analyzed interferometric CS $J$=2--1 line maps of the Horsehead PDR 
at a $3.65'' \times 3.34''$  resolution together with single--dish observations of several
rotational lines of CS, C$^{34}$S and HCS$^+$. We have studied the CS photochemistry,
excitation and radiative transfer using the latest  HCS$^+$ and OCS$^+$ dissociative
recombination rates  (Montaigne et al. 2005) and CS collisional cross--sections 
(Lique et al. 2006). The main conclusions are as follows (the
whole work can be found in Goicoechea et al. (2006)):

\begin{enumerate}

\item CS and C$^{34}$S rotational line emission reveals
mean densities around $n(H_2$)=(0.5--1.0)$\times$10$^{5}$~cm$^{-3}$.
The CS $J$=5--4 lines show narrower line widths than the low--$J$ CS lines  and  
require higher density gas components, $\sim$(2--6)$\times$10$^{5}$~cm$^{-3}$, not  
resolved by  a  $\sim$10$''$ beam.
These values are larger than previous estimates based on CO single--dish observations.
It is likely that  clumpiness at scales below $\sim$0.01~pc and/or a low density 
envelope play a role in the CS line profile formation.

\item Nonlocal, non--LTE radiative transfer models of optically thick CS lines
and optically thin C$^{34}$S lines provide an accurate determination
of the CS abundance, $\chi$(CS)=(7$\pm$3)$\times$10$^{-9}$.
We show that radiative transfer and opacity effects play a role
in the resulting CS line profiles but not in C$^{34}$S lines. 
Assuming the same physical conditions for the HCS$^+$ molecular ion,
 we find $\chi$(HCS$^{+}$)=(4$\pm$2)$\times$10$^{-11}$.

\begin{figure*}[t]
  \includegraphics[width=0.3\textwidth, angle=-90]{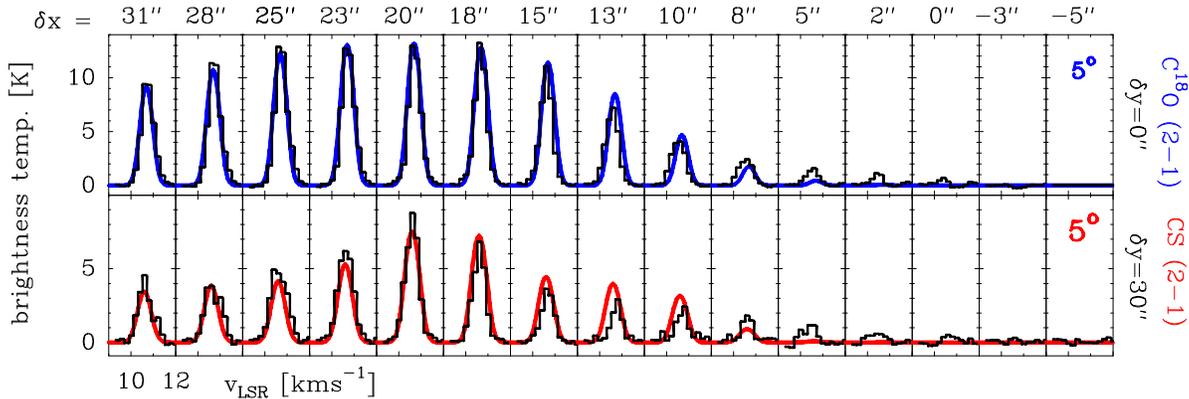}
\caption{PdBI C$^{18}$O $J$=2--1 and CS $J$=2--1 spectra along
the exciting star direction,  $\delta y = 0''$ (\textit{upper panel})
and $\delta y = 30''$ (\textit{lower panel}). 
Monte Carlo radiative transfer models using the output of PDR models 
for C$^{18}$O (blue curve) and  CS (red curve) for the physical conditions 
discussed in the text (assuming that the PDR
is inclined relative to the line of sight by a $\varphi$=5$^o$ angle).
Modeled line profiles have been convolved with an appropriate gaussian
beam corresponding to each synthesized beam.}
\label{fig:3}       
\end{figure*}

\item According to photochemical models, the gas phase sulfur abundance 
required to reproduce these CS and HCS$^+$ abundances is 
S/H=(3.5$\pm$1.5)$\times$10$^{-6}$, only a factor $\sim$4 less abundant than the
solar elemental abundance.
Larger sulfur  abundances are possible if the gas is significantly warmer.
Thus, the sulfur abundance in the PDR is very close to the
undepleted value observed in the diffuse ISM.
The predicted CS/HCS$^+$ abundance ratio is  close to the observed
value of $\sim$175, especially if predicted HCS$^+$ peak abundances
are considered. If not,  the HCS$^+$ production is  underestimated unless 
the gas is in a higher ionization phase, e.g.
if the cosmic ray ionization rate is increased by  $\sim$5.
We are currently working on a larger inventory of sulfur--bearing molecules detected 
by us in the Horsehead PDR 
to test other limitations of the sulfur chemical network.

\item High angular resolution PdBI maps reveal that the CS  emission
does not follow the same morphology shown by the small hydrocarbons emission
the PDR edge (Pety et al. 2005).  In combination with 
previous PdBI C$^{18}$O  observations we have modeled the PDR edge and confirmed that a
 steep density gradient (see Habart et al. 2005) is needed to reproduce CS and C$^{18}$O
observations.  The resulting density profile qualitatively agrees to that predicted  in 
numerical simulations of a shock front compressing the PDR edge to high 
densities, $n(H_2$)$\simeq$10$^{5}$~cm$^{-3}$, and
high thermal pressures, $\simeq$(5--10)$\times$10$^6$~K~cm$^{-3}$.

\item Conventional PDR heating and cooling mechanisms fail to reproduce
the temperature of the warm gas observed in the region  by at
least a factor $\sim$2. Additional mechanical heating mechanisms associated with the gas 
dynamics may be needed to account for the warm gas. The thermal structure of the PDR
edge is still not fully constrained from observations. This fact
adds uncertainty to the abundances predicted by photochemical models.

\end{enumerate}

Observational studies of ISM clouds show that many physical and chemical variations
occur at very small angular scales. For PDRs, an accurate molecular
inventory, as a function of the distance from the illuminating source,
can only be obtained from interferometric observations.
Aperture synthesis observations contain detailed information
about density, temperature, abundance and structure of the cloud,
but only specific radiative transfer and photochemical models for each
given source are able to extract
the information. A minimum description of the source geometry is
usually needed. Future observations with ALMA will allow us to
finally  characterize PDRs at the appropriate spatial scales
of their physical and chemical gradients.\\

\begin{acknowledgements}
J.R. Goicoechea was supported by the French \textit{Direction de la Recherche} 
and by a \textit{Marie Curie Intra-European Individual
Fellowship} within the 6th European Community Framework Programme,
contract MEIF-CT-2005-515340.
\end{acknowledgements}



\end{document}